\documentclass[twocolumn,showpacs,prc,superscriptaddress,proprietresses]{revtex4}
\usepackage[dvips]{graphicx}
\usepackage{amsmath,amssymb}
\usepackage{dcolumn}
\begin{document}
\title{Multiplicity fluctuations due to the temperature fluctuations
in high-energy nuclear collisions }
\author{Grzegorz Wilk}
\email{wilk@fuw.edu.pl}
\affiliation{The Andrzej So{\l}tan Institute for Nuclear Studies,
Ho\.{z}a 69, 00681, Warsaw, Poland}
\author{Zbigniew W\l odarczyk}
\email{wlod@pu.kielce.pl} \affiliation{Institute of Physics,
                Jan Kochanowski University,  \'Swi\c{e}tokrzyska 15,
                25-406 Kielce, Poland}
\date{\today}
\begin{abstract}
We investigate the multiplicity fluctuations observed in
high-energy nuclear collisions attributing them to intrinsic
fluctuations of temperature of the hadronizing system formed in
such processes. To account for these  fluctuations we replace the
usual Boltzmann-Gibbs (BG) statistics by the non-extensive Tsallis
statistics characterized by the nonextensivity parameter $q$, with
$|q-1|$ being a direct measure of fluctuations. In the limit of
vanishing fluctuations, $q \rightarrow 1$ and Tsallis statistics
converges to the usual BG. We evaluate the nonextensivity
parameter $q$ and its dependence on the hadronizing system size
from the experimentally observed collision centrality dependence
of the mean multiplicity, $\langle N\rangle$, and its variance,
$Var(N)$. We attribute the observed system size dependence of $q$
to the finiteness of the hadronizing source with $q = 1$
corresponding to an infinite, thermalized source with a fixed
temperature, and with $q > 1$ (which is observed) corresponding to
a finite source in which both the temperature and energy
fluctuate.

\end{abstract}

\maketitle

\section{\label{sec:I}Introduction}

With the large number of particles produced in heavy ion
collisions at CERN SPS and BNL RHIC it is possible to study
fluctuations in different physical observables on the
event-by-event basis \cite{HH}. These fluctuations are potentially
a very important source of information on the thermodynamical
properties of strongly interacting systems formed in such
collisions, like, for example, its specific heat \cite{SS}
(connected with fluctuations observed in particle multiplicities
\cite{NA49,WA98}, in transverse momenta \cite{NA49pT} and in other
global observables), its chemical potential or matter
compressibility \cite{StM}.

Fluctuations of multiplicity observed in heavy ion collisions
exhibit spectacular and unexpected features as functions of the
number of participants. Recent results on centrality dependence in
$Pb+Pb$ collisions at $158$A GeV  obtained by NA49 \cite{NA49} and
WA98 \cite{WA98} experiments indicate that the scaled variance of
the multiplicity distribution, $Var(N)/\langle N\rangle$,
increases when proceeding from the central towards peripheral
collisions, i.e., when the number of participants decreases. Such
behavior is confirmed by a comprehensive survey of multiplicity
fluctuations of charged hadrons provided by PHENIX experiment
\cite{PHENIX}.

During the last decade the statistical models of strong
interactions were constantly used as an important tool to study
fluctuation pattern observed in high energy nuclear collisions
experiments, which are quantified by scaled variance mentioned
above (see, for example, \cite{MIG1,MIG2,MIG3} and references
therein). However, to minimize the effect of the possible
participant number fluctuations, the analysis of particle number
fluctuations has been restricted to the most central $A+A$
collisions only.

However, up to now, none of the models aimed to describe all
essential features of multiparticle production processes (like
multiplicities and distributions of particles in phase space and
their composition) describe also the dependence of $Var(N)/\langle
N\rangle$ on $N_P$, the number of nucleons from projectile nucleus
participating in the collision - projectile participant, observed
experimentally \cite{NA49}. They lead to multiplicity
distributions of the (approximately) Poissonian form, independent
of centrality, which highly underestimate the observed
multiplicity fluctuations in noncentral collisions and thus are
unable to reproduce, even qualitatively, the observed centrality
dependence of the scaled variance. This remark applies not only to
the Monte Carlo models like HIJING \cite{HIJING}, HSD \cite{KLGB}
or UrQMD \cite{UrQMD}, which are based on string excitation and
decay, but also to any statistical model that does not assume
correlations among secondary particles (cf. discussion in Sec.
\ref{sec:IIIC}). On the other hand, these results can be described
by some specialized models addressing fluctuations directly, like
the percolation model \cite{percolation}, the model assuming
inter-particle correlations caused by the combination of strong
and electromagnetic interactions \cite{MW} or the transparency,
mixing and reflection model \cite{MixRefl}. Fluctuations in these
models reflect some dynamical features of the production process,
(specific for the model considered).

In this paper we shall address the problem of multiplicity
fluctuations without resorting to any specific dynamical picture
but, instead, by attributing them to some nonstatistical,
intrinsic fluctuations existing in a hadronizing system produced
in high energy heavy ion collisions. To account for such
fluctuations we shall use a special version of statistical model
based on nonextensive Tsallis statistics \cite{Tsallis} in which
fluctuations of the temperature are known to be directly connected
with the nonextensivity parameter $q$ \cite{WW,BiroJ}, namely $ q
= 1 + Var(1/T)/\langle 1/T\rangle ^2 $. In what follows they will
be assumed to be the true (if not the only) origin of the
fluctuations observed in the experimental data. The resulting
distributions are then power-like, $\exp( - E/T) \Longrightarrow
\exp_q(-E/T) = [1 - (1 - q)E/T]^{1/(1-q)}$ and in the limit of $ q
\rightarrow 1 $ they go smoothly to the usual Boltzmann
distributions (in what follows we shall address only the $q \ge 1$
case) \cite{FOOT3}. It is important for our further discussion
that such non-exponential distributions of energy result in a
non-Poissonian multiplicity distributions of the produced
secondaries \cite{WWfluct}.

However, to incorporate the new features of data reported by
\cite{DataPi}, we have to extend the notion of fluctuating
temperature replacing it by some $q$-dependent {\it effective
temperature}, $T_{eff}$, which accounts not only for the intrinsic
fluctuations in the hadronizing source (as in \cite{WW}) but also
for effects of the possible energy transfer taking place between
the hadronizing source and its surroundings  \cite{WW_epja}. This
will be done in Section \ref{sec:II}. Section \ref{sec:III}
contains our results: the universal participant dependence (i.e.,
scaling in the variable $ f = N_P/A$) presented in Section
\ref{sec:IIIA}, its explanation by the observation that in reality
the hadronizing source is always of finite size presented in
Section \ref{sec:IIIB} and the system size dependence of
multiplicity and multiplicity fluctuations presented in Sections
\ref{sec:IIIC}. Section \ref{sec:IV} summarizes our work. Some
details are presented in Appendices \ref{sec:Teff_derivation} and
\ref{sec:P(N)_derivation}.

\section{\label{sec:II}Effective temperature}

In the proposed approach we replace the standard Boltzmann-Gibbs
exponential distribution,
\begin{equation}
g(E) = C\exp( - E/T), \label{eq:exp}
\end{equation}
by the Tsallis distribution ($q$-exponential) defined as
\begin{equation}
h_q(E) = C_q\left[ 1 -
(1-q)\frac{E}{T_{eff}}\right]^{\frac{1}{1-q}}, \label{eq:q_exp}
\end{equation}
where
\begin{equation}
q = 1 + \frac{Var(T)}{\langle T\rangle^2}. \label{eq:defq}
\end{equation}
 \begin{figure}[t]
  \begin{center}
   \includegraphics[width=9cm]{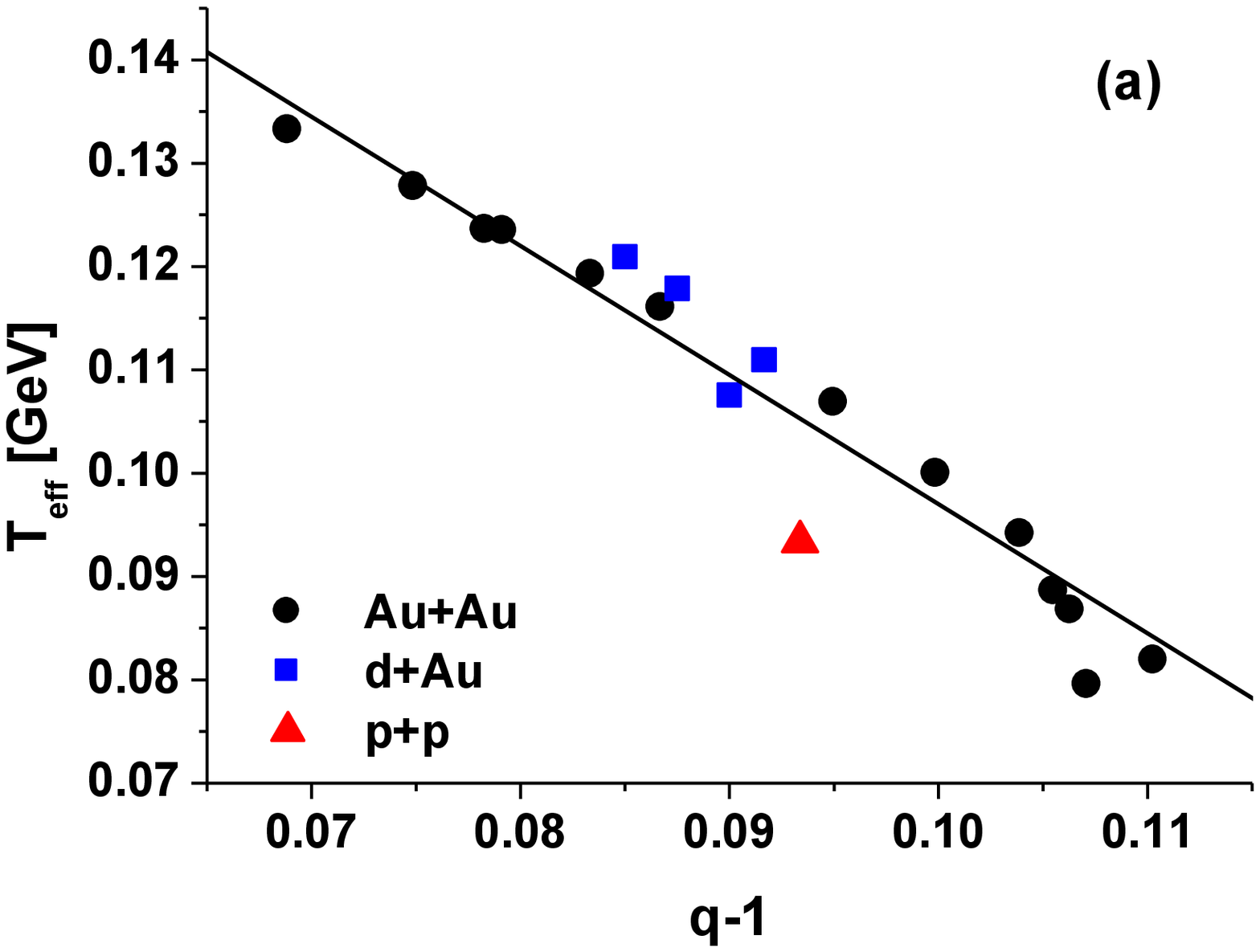}\\
   \vspace{-7.5cm}
   \includegraphics[width=9cm]{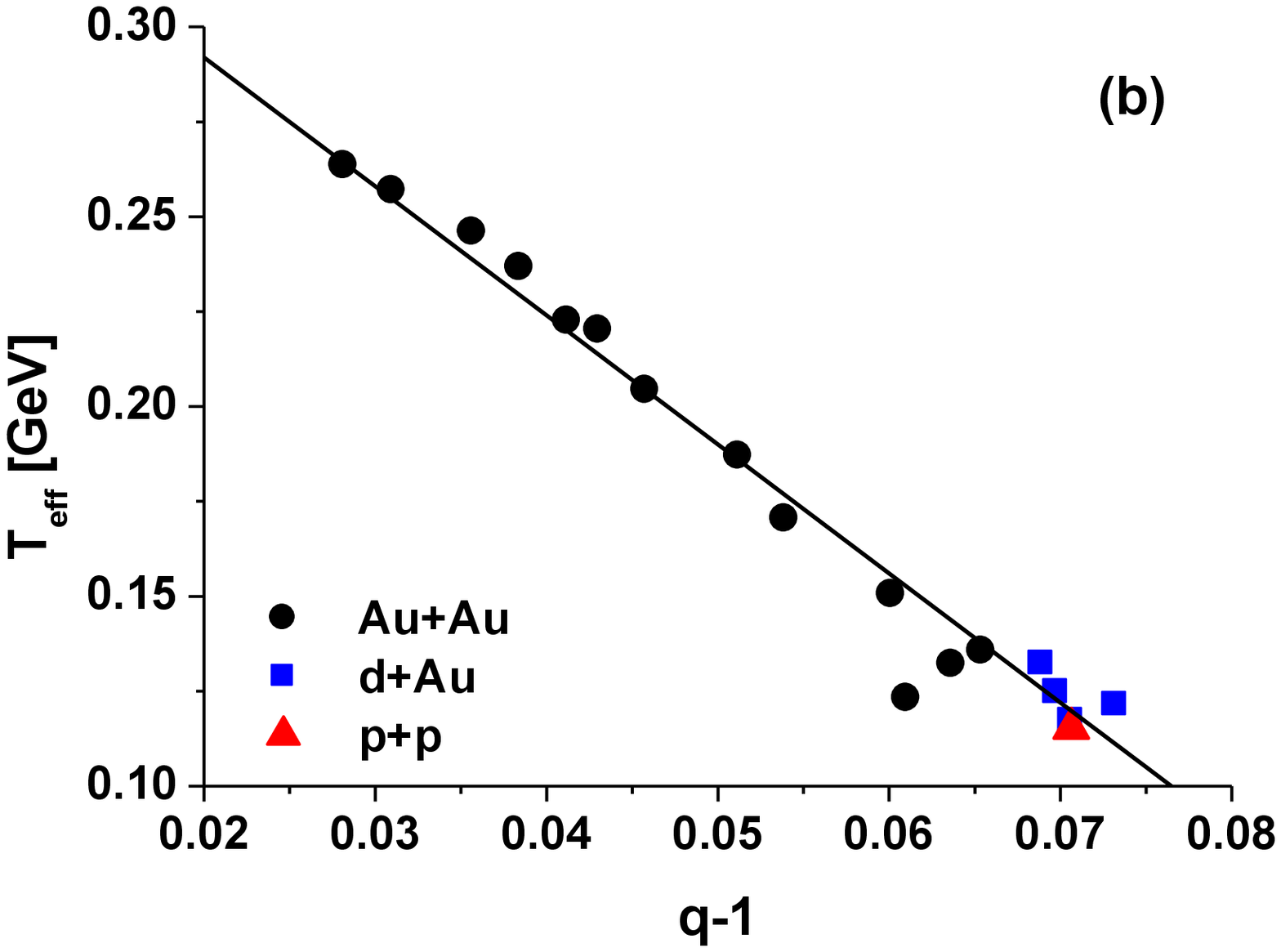}
   \vspace{-6.5cm}
   \caption{(Color on line) Dependence of the effective temperature $T_{eff}$ (in GeV)
            on the parameter $q$ for the production in different reactions
            of negative pions $(a)$  and antiprotons $(b)$,
            all data points are from \cite{DataPi}. The solid lines show linear
            fits to the obtained results: for upper panel $T_{eff}
            = 0.22 - 1.25 (q-1)$, for lower panel $T_{eff} = 0.36 - 3.4 (q-1)$.}
   \label{Fig:1}
  \end{center}
\end{figure}
and
\begin{equation}
T_{eff} = T_0 + (q - 1) \frac{1}{Dc_p\rho}\phi, \label{eq:Teff}
\end{equation}
with $c_p$, $\rho$ and $D$ being, respectively, the specific heat
under constant pressure, density and the strength of the
temperature fluctuations (cf. Appendix \ref{sec:Teff_derivation}
for details). Effective temperature  $T_{eff}$ occurs when we
experience both the fluctuations of the temperature $T$ (around
the value $T_0$) and some energy transfer taking place between the
source and the surroundings given by $\phi$ \cite{FOOT4}. Notice
that in Eq. (\ref{eq:Teff}) energy transfer affects $T_{eff}$ only
in the presence of fluctuations, i.e., for $q \neq 1$. The
predicted $q$ dependence of $T_{eff}$ is indeed observed
experimentally, cf. Fig. \ref{Fig:1}. Namely, in \cite{DataPi} the
transverse momentum spectra of pions and antiprotons produced in
the interactions of p+p, d+Au and Au+Au at $\sqrt{s_{NN}}=200$ GeV
at RHIC-BNL experiments \cite{DataPantiP} were analyzed using a
nonextensive approach in which the slope of $p_T$ distribution
determines the effective temperature $T_{eff}$. Its shape gives
the parameter of nonextensivity $q$ \cite{FOOT5}. They evaluated,
among other things, the nonextensivity parameter $q$ and the
effective temperature $T_{eff}$ for a different number of
participants, $N_P$. From them we can deduce the dependence of
$T_{eff}$ on the parameter $q$ which is shown in Fig. \ref{Fig:1}.
Notice that in all cases we find that $\phi < 0$ and that
$T_{eff}$ seems to depend linearly on $(q-1)$. Negative values of
$\phi$ mean that the energy is transferred from the interaction
region to the surroundings (i.e., to the spectators of
noninteracting nucleons) \cite{FOOT6}.

\section{\label{sec:III}Different facets of multiplicity fluctuations}

\subsection{\label{sec:IIIA}The universal participant dependence}

We start a discussion of multiplicity fluctuations by recollecting
the result of \cite{WWfluct} saying that if $N$ particles are
distributed in energy according to the $N$-particle Tsallis
distribution described by the nonextensive parameter $q$ then
their multiplicity distribution {\it has to be} of the
Negative-Binomial type with $k^{-1} = q - 1$ (cf. Appendix
\ref{sec:P(N)_derivation} for details). It means therefore that we
can expect that
\begin{equation}
\frac{\frac{Var(N)}{\langle N\rangle} - 1}{\langle N\rangle} = q -
1 \label{eq:der1}
\end{equation}
(and, what is important, that $q - 1$ determined this way does not
depend on the acceptance of the detector \cite{FOOT-I}). On the
other hand, if $U$ is the accessible energy and $gT_{eff}$ is the
mean energy per particle detected in the acceptance region (with
$g$ being a parameter and $T_{eff}$ effective temperature defined
in Eq. (\ref{eq:Teff})), than
\begin{equation}
\langle N\rangle = \frac{\langle U\rangle}{gT_{eff}}. \label{eq:U}
\end{equation}
Using Eq. (\ref{eq:Teff}) one can shown that
\begin{equation}
\frac{\langle N\rangle - n_0 N_P}{\langle N\rangle} = c (q - 1),
\label{eq:der2}
\end{equation}
where $n_0$ is the multiplicity in the single nucleon-nucleon
collision measured in the region of acceptance, it is defined by
the constraint that $\langle U\rangle = n_0 gT_0 N_P$, whereas $c
= -\phi/\left( Dc_p\rho T_0 \right)$ is a constant (notice that
because in cases we are interested here $\phi < 0$, cf. Fig.
\ref{Fig:1}, $c$ is positive). Comparing now Eqs. (\ref{eq:der1})
and (\ref{eq:der2}) one can expect that
\begin{equation}
\frac{Var(N)}{\langle N\rangle} = 1 + c\left( \langle N\rangle -
n_0 N_P\right). \label{eq:comp}
\end{equation}
Notice that Eqs. (\ref{eq:der2}) and (\ref{eq:comp}) do not depend
on parameter $g$ anymore; its role in present work is to fit, if
necessary, $\langle N\rangle$ in Eq. (\ref{eq:U}) to experimental
data only.

\begin{figure}[t]
  \begin{center}
   \includegraphics[width=9cm]{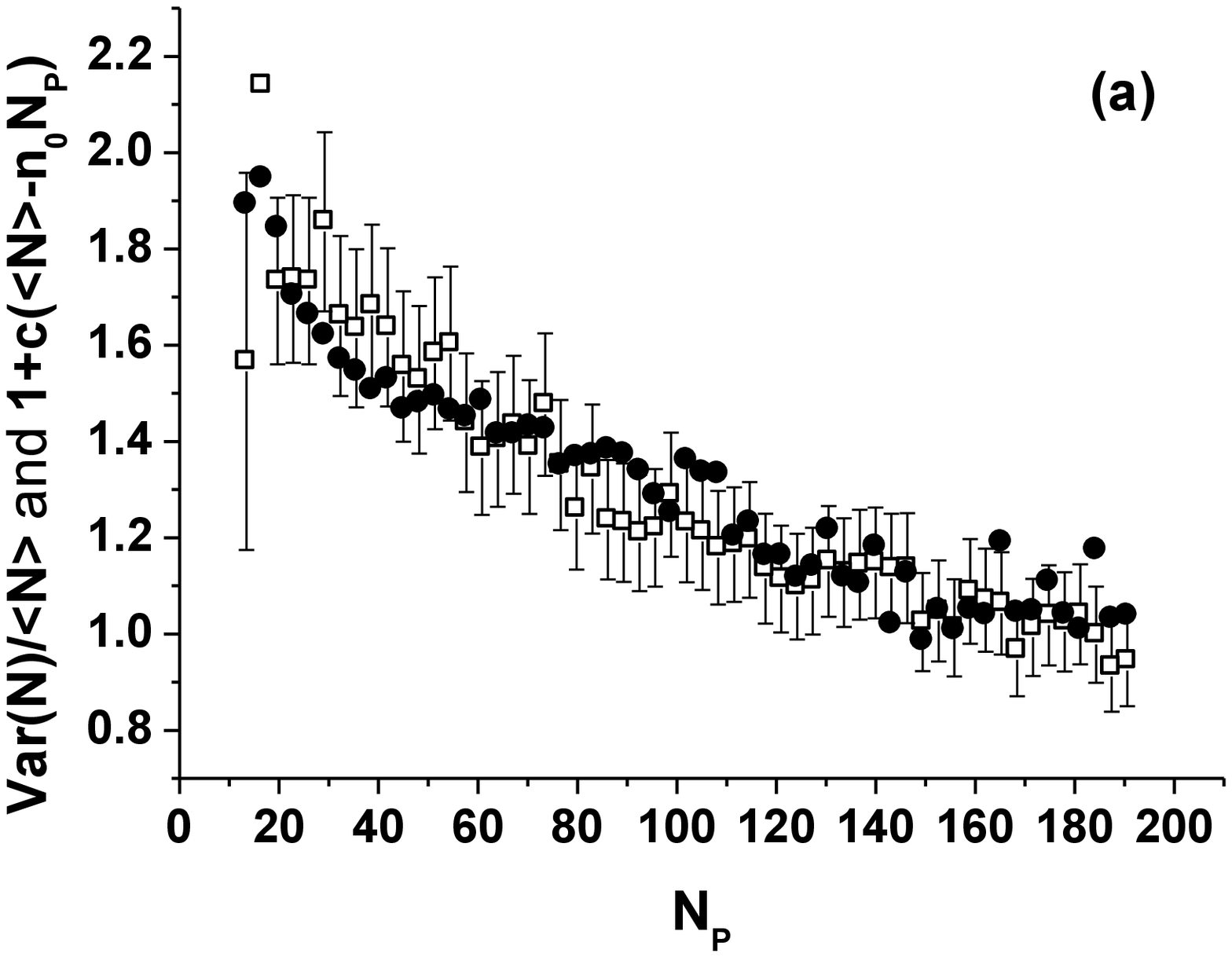}\\
   \vspace{-7.5cm}
   \includegraphics[width=9cm]{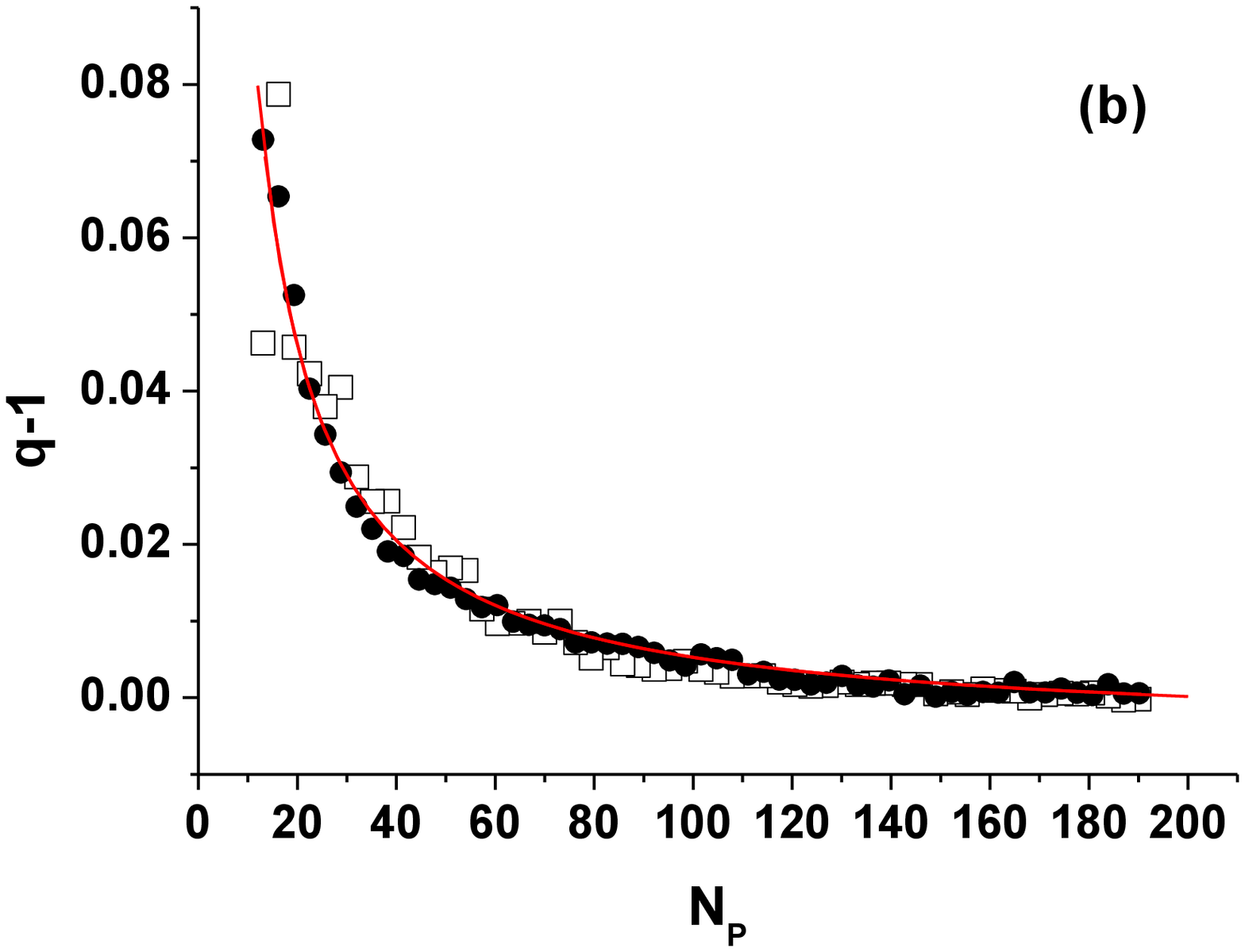}
   \vspace{-6.5cm}
   \caption{(Color on line) $(a)$ Comparison of $Var(N)/\langle N\rangle$ versus $N_P$ (squares)
            with $1 + c (\langle N\rangle - n_0N_P)$ versus $N_P$
            (circles) for $n_0 = 0.642$ and $c=4.1$ \cite{FOOTA}. Data  are for
            negatively charged particles from $Pb+Pb$ collisions
            as collected by NA49 experiment \cite{NA49}. $(b)$ The same but translated to dependence of $q-1$
            versus $N_P$. Squares were obtained from the $Var(N)/\langle N\rangle$
            versus $N_P$ dependence whereas circles from the
            $\langle N\rangle$ versus $N_P$ dependence. The values
            of $n_0$ and $c$ are the same as before. The solid
            line shows the dependence (\ref{eq:q-1}) for the
            nuclear mass number $A=207$ and parameter $a=0.98$.}
   \label{Fig:2}
  \end{center}
 \end{figure}

This means that using the notion of $T_{eff}$ one should observe
that $Var(N)/\langle N\rangle$ and $\langle N\rangle$ are mutually
connected. Fig. \ref{Fig:2} shows that this is indeed the case. If
$T_{eff}$ depends on $q$ (i.e., for $ c\neq 0$) than the
dependence $Var(N)/\langle N\rangle$ on the number of participants
$N_P$ is connected to the dependence of $\langle N\rangle $ on
$N_P$. Notice that if $\langle N\rangle $ is linear in $N_P$ then
$Var(N)/\langle N\rangle$ is constant. In particular, for $\langle
N\rangle = n_0N_P$ we have $Var(N)/\langle N\rangle = 1$.
Therefore the experimental fact that $Var(N)/\langle N\rangle$
decreases with increasing $N_P$ indicates {\it nonlinear}
dependence of  $\langle N\rangle$ on the number of participants
$N_P$.

It should be stressed that data taken by the NA49 experiment
\cite{NA49} show that $Var(N)/\langle N\rangle$ changes rather
strongly with $N_P$. For peripheral collisions (i.e., for small
$N_P$) one observes marked deviation of  $Var(N)/\langle N\rangle
$ from unity. At the same time deviation of $\langle N\rangle$
from linearity in $N_P$ is very weak. Our result based on the
concept of effective temperature, $T_{eff}$, and showing that
$Var(N)/\langle N\rangle$ and $\langle N\rangle$ are connected is
therefore by no means trivial.

We close this Section by reminding that all previous attempts in
this field were addressing only the behavior of $Var(N)/\langle
N\rangle$ versus $N_P$, leaving aside the possible nonlinear
dependence of $\langle N\rangle$ on $N_P$. Fig. \ref{Fig:2} shows
that introducing $T_{eff}$ results in a kind of scaling, namely
the same value of $q - 1$ describes dependencies of both variance
and mean multiplicity on $N_P$, following Eq. (\ref{eq:comp}),
which - we repeat - has its origin in the $q$-dependence of
$T_{eff}$ \cite{FOOT7}.

\subsection{\label{sec:IIIB}Finite hadronizing source size and temperature
fluctuations}

We shall now derive the $q - 1$ dependence on $N_P$ seen in Fig.
\ref{Fig:2}. Let us first observe that \cite{LL1,SS}
\begin{equation}
q - 1 = \frac{Var(T)}{\langle T\rangle ^2} = \frac{1}{C_V},
\label{eq:CV}
\end{equation}
i.e., the parameter $q$ can be regarded as connected (via
fluctuations of temperature) to the heat capacity under constant
volume, $C_V$. For a system with finite size remaining in contact
with a heat bath one has, following Lindhard's approach
\cite{Lin}, that
\begin{equation}
Var(U) + C_V^2 Var(T) = \langle T\rangle ^2 C_V, \label{eq:unrel}
\end{equation}
This is a kind of uncertainty relation (in the sense that it
expresses the truth that in the case of conjugate variables one
standard deviation in some measurement can only become small at
the expense of the increase of some other standard deviation
\cite{UnRel}). Relation (\ref{eq:unrel}) is supposed to be valid
all the way from the canonical ensemble, where $Var(T) = 0$ and
$Var(U) = \langle T\rangle ^2 C_V$, to the microcanonical
ensemble, for which $Var(T) = \langle T\rangle ^2/C_V$ and $
Var(U) = 0$. Eq. (\ref{eq:unrel}) expresses both the
complementarity between the temperature and energy, and between
the canonical and microcanonical description of the system
\cite{FOOT8}.

To obtain realistic (intermediate) distributions, start from a
system at a fixed temperature $T$. The standard deviation of its
energy is
\begin{equation}
 Var(U) = \langle T\rangle^2
\frac{\partial \langle U\rangle}{\partial T} = \langle T\rangle ^2
C_V. \label{eq:eq1}
\end{equation}
Inverting the canonical distribution $g_T(U)$ one can obtain
\begin{equation} g_U(T) = -T^2 \frac{\partial}{\partial T}
\int^U_0g_T(U')dU' \label{eq:invert}
\end{equation}
and interpret it as a probability distribution of the temperature
in the system (cf., Eq. (\ref{eq:Gamma})). The standard deviation
of this distribution  then yields \cite{FOOT9}
\begin{equation}
Var(T) = \frac{\langle T\rangle ^2}{C_V}.
\label{eq:microcanonical}
\end{equation}
Because for a canonically distributed system the energy variance
is $ Var(U) = \langle T\rangle^2 C_V$ and because for an isolated
system $Var(U) = 0$, for the intermediate case the variance
(expressing energy fluctuations in the system) can be assumed to
be equal to
\begin{equation}
Var(U) = \langle T\rangle^2 C_V \xi,\quad \xi \in (0,1),
\label{eq:eq2}
\end{equation}
where the parameter $\xi$ depends on the size of the hadronizing
source. Inserting this in Eq. (\ref{eq:unrel}) one gets that $q$
depends on $\xi$ in the following way:
\begin{equation}
q - 1 = \frac{Var(T)}{\langle T\rangle^2} = \frac{1 - \xi}{C_V}.
\label{eq:relation}
\end{equation}
Assuming now that the size of the thermal system produced in heavy
ion collisions is proportional to the number of participating
nucleons $N_P$, i.e., that $\xi \simeq f = N_P/A$,  and taking
into account that $C_V \cong a N_P $, we obtain that
\begin{equation} q - 1 = \frac{1}{a N_P}( 1- f ). \label{eq:q-1}
\end{equation}
This is the relation which nicely fits data, cf., Fig.
\ref{Fig:2}.

To recapitulate: we know that if $U =$ const and $T =$ const then
the multiplicity distribution $P(N)$ is Poissonian (cf. Appendix
\ref{sec:Poisson}) and we also know how fluctuations of $T$ change
$P(N)$ from the Poissonian form to the NBD one (cf. Appendix
\ref{sec:NBD}). We want now to see how big are fluctuations of $T$
in our hadronizing systems formed in a collision process. It turns
out that for $N =$ const we have either Eq. (\ref{eq:eq1}) or Eq.
(\ref{eq:microcanonical}) depending on whether $T =$ const or $U
=$ const. We claim then that we can assume validity of Eq.
(\ref{eq:unrel}) not only in the above limiting cases but also in
the general case when both the energy $U$ and the temperature $T$
fluctuate at the same time, i.e., if fluctuations of the energy
$U$ are given by Eq. (\ref{eq:eq2}) then fluctuations of the
temperature $T$ are given by Eq. (\ref{eq:relation}). Finally,
knowing how big are fluctuations of $T$ we can deduce fluctuations
of the multiplicity $N$. This will be performed in what follows.\\

\subsection{\label{sec:IIIC}System size dependence of mean
multiplicity and multiplicity fluctuations}

\begin{figure}[h]
\vspace{-1cm}
  \begin{center}
   \includegraphics[width=11.0cm]{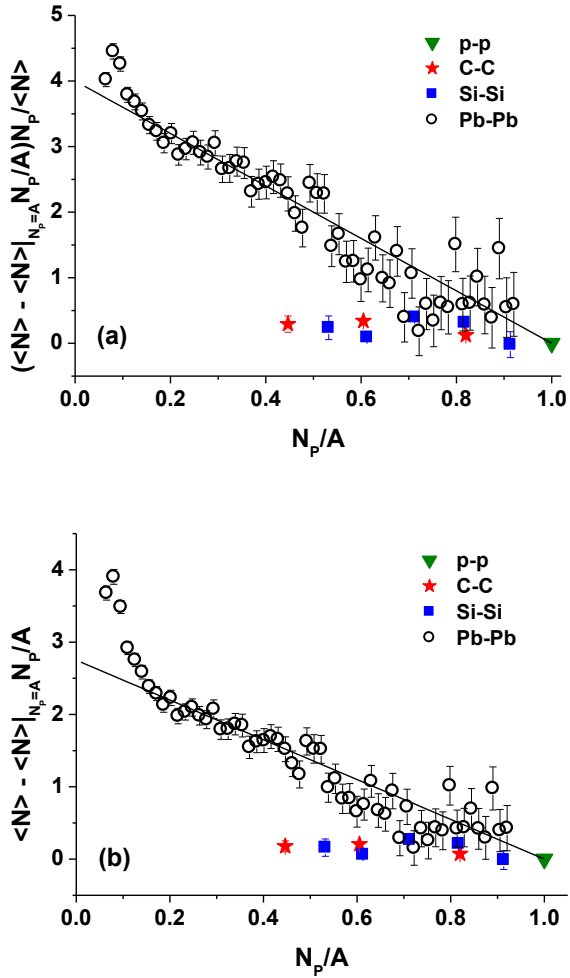}
   \vspace{-1.5cm}
   \caption{(Color online) Mean multiplicities compared with predictions given
            by Eq.(\ref{eq:prediction1}) $(a)$
            and Eq.(\ref{eq:prediction2}) $(b)$.
            We adopt $n_0 = 0.559$, $0.575$, $0.657$ and $0.642$
            for, respectively, $p+p$, $C+C$, $Si+Si$ and $Pb+Pb$ collisions.
            Linear fit show the predictions of Eq. (\ref{eq:prediction2}) with
            parameter $c/a' = 2.75$  and Eq. (\ref{eq:prediction1}) with parameter $c/a = 1.8$.
            Only statistical errors are indicated.}
   \label{Fig:5}
 \end{center}
\end{figure}

We shall now discuss, respectively, the system size dependence of
the mean multiplicity and multiplicity fluctuations. In our
approach the system size enters through $C_V$ and $\xi$. We shall
keep $\xi = N_P/A$ and connect $C_V$ with the measured quantities
considering two natural assumptions: either
\begin{equation}
C_V = a N_P \label{eq:a}
\end{equation}
or
\begin{equation}
C_V = a' \langle N\rangle .\label{eq:aprim}
\end{equation}
When used together with Eqs.(\ref{eq:der2}) and
(\ref{eq:relation}) the second possibility leads to a very simple
scaling relation:
\begin{equation}
\langle N\rangle - \frac{N_P}{A}\langle N\rangle|_{N_P=A} =
\frac{c}{a'}\left( 1 - \frac{N_P}{A}\right) \label{eq:prediction2}
\end{equation}
whereas the first one results in a slightly more involved formula
\begin{equation}
\left(\langle N\rangle - \frac{N_P}{A}\langle N\rangle |_{N_P=A}
\right)\frac{N_P}{\langle N\rangle} = \frac{c}{a}\left( 1 -
\frac{N_P}{A}\right). \label{eq:prediction1}
\end{equation}
In both cases $\langle N\rangle|_{N_P = A} = n_0 A$ is
multiplicity extrapolated to $N_P = A$. As seen in Fig.
\ref{Fig:5}, where comparison with experimental data \cite{NA49}
is presented, one observes different dependencies for $Pb+Pb$
collision and lighter nuclei for which in semicentral collisions
such an effect is practically not observed. In addition, for
peripheral collisions ($N_P < 0.15 A$ ) one observes the deviation
from the expected dependence (\ref{eq:prediction2}). Notice that
deviation from the linear fit for $Pb+Pb$ collisions concern only
the $5$ most peripheral points and the observed discrepancy means
that the measured experimentally mean multiplicity is less than
$1$ particle higher than the expected value. The agreement with
prediction given by Eq. (\ref{eq:prediction1}) seems to be better,
what means that the assumption $C_V = aN_P$ is more realistic.

\begin{figure}[h]
 \vspace{-1cm}
  \begin{center}
   \includegraphics[width=9.0cm]{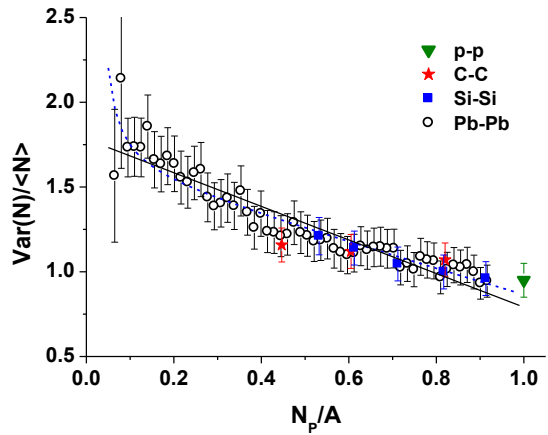}
   \vspace{-6.5cm}
   \caption{(Color online) The scaled variance of the multiplicity distribution of negatively
   charged particles produced in p+p, semicentral C+C, semicentral Si+Si, and
   Pb+Pb collisions as function of the fraction of nucleons participating in
   the collision, $N_P/A$. The experimental data \cite{NA49} are compared with
   prediction (\ref{eq:eq4}) with parameters $\omega_0 = 0.79$ and
   $a' = 1.0$ (solid line) and with prediction (\ref{eq:22}) with parameters
   $\omega_0 = 0.87$ and $a = 0.83$ (dashed line, the other parameters are the same as in
   Fig. \ref{Fig:2}  with $n_0 = 0.642$ and $c = 4.1$ for $A = 207$). }
   \label{Fig:3}
  \end{center}
  \vspace{-0.5cm}
\end{figure}

In what concerns multiplicity fluctuations, assumption
(\ref{eq:aprim}) results (when combining Eqs. (\ref{eq:q-1}) and
(\ref{eq:der1})) in the following simple scaling relation:
\begin{equation}
\frac{Var(N)}{\langle N\rangle} = 1 + \langle N\rangle (q - 1) = 1
+ \frac{1}{a'}\left( 1 - \frac{N_P}{A} \right), \label{eq:eq3}
\end{equation}
i.e., in the multiplicity fluctuations dependent on the fraction
of nucleons participating in the collision $f = N_P/A$. Such
scaling was found recently by the NA49 collaboration \cite{NA49}
when comparing collisions of $C+C$, $Si+Si$ and $Pb+Pb$, cf. Fig.
\ref{Fig:3}. If, instead, we use (\ref{eq:a}) then, taking into
account dependence of $\langle N\rangle$ on $N_P$ given by Eq.
(\ref{eq:der2}), one obtains that, more exactly,
\begin{equation}
\frac{Var(N)}{\langle N\rangle} = 1 + \langle N\rangle (q - 1) = 1
+ \frac{n_0\left( 1 - \frac{N_P}{A} \right)}{a - \frac{c}{A}\left(
\frac{A}{N_P} - 1\right)}. \label{eq:22}
\end{equation}

We observe thus a weak dependence on the mass number $A$ of
colliding nuclei. For a small number of participants, $N_P << A$,
we observe an additional increase of relative variance,
$Var(N)/\langle N\rangle$, in comparison with prediction
(\ref{eq:eq3}). Notice that for $\phi > 0$ one expects just the
opposite trend, i.e., the non-monotonic behavior of
$Var(N)/\langle N\rangle$ with increasing number of participants,
$N_P$. The question about the sign of the heat source term $\phi$
seems to still be open due to the lack of data in the region of
small number of participants.

For central collisions (i.e., for $f = N_P/A \simeq 1$)
multiplicity distributions are sub-Poissonian and $[Var(N)/\langle
N\rangle]|_{N_P = A} = \omega_0 < 1$. For example, within a
statistical model with fixed volume $\omega_0$ varies in the range
$0.5 \div 1.0$ \cite{BGGZ}. In particular (see \cite{MIG1,MIG3}
and references therein): $(i)$ The global conservation laws
imposed on each microscopic state of the statistical system lead
to suppression of the particle number fluctuations. The final
state scaled variance behavior in the canonical ensemble is
characterized by $\omega_0 \cong 0.8$ whereas in the
microcanonical ensemble by $\omega_0 \cong 0.3$. $(ii)$  In the
primordial values of the scaled variance (i.e., calculated before
decays of resonances) one can also observe the effect of quantum
statistics. It turns out that proper inclusion of Bose statistics
leads to the additional enhancement of particle number
fluctuations of the order of $\omega_0 \cong 1.05 \div 1.06$,
which is quite small at the chemical freeze-out.  $(iii)$ The
particle number fluctuations can be also enhanced by including
explicite production and decay of resonances (either in the grand
canonical ensemble of in canonical ensemble). They lead to
$\omega_0 \cong 1.1$. The actual value of parameter $\omega_0$
depends on the energy of collision and on the acceptance in which
the multiplicity fluctuations are observed and varies in the range
$0.8 - 1.0$ \cite{CAlt}. For this reason in Fig. \ref{Fig:3} we
compare experimental data with the formula
\begin{equation}
\frac{Var(N)}{\langle N\rangle} = \omega_0 + \frac{1}{a'}\left( 1
- \frac{N_P}{A}\right).  \label{eq:eq4}
\end{equation}

\begin{figure}[h]
 \vspace{-1cm}
  \begin{center}
   \includegraphics[width=9.0cm]{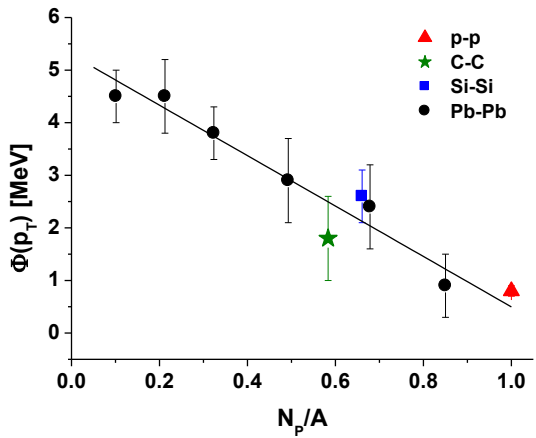}
   \vspace{-6.5cm}
   \caption{(Color online) Transverse momentum fluctuations of negative particles defined
            by the measure $\Phi$ \cite{PHI} as a function
            of number of the fraction of nucleons participating in the collision,
            $N_P/A$. The triangles correspond to $p+p$ collisions, asterisks to
            $C+C$, squares to $Si+Si$, and circles to $Pb+Pb$.
            The experimental data are taken from \cite{NA49pT} and are compared
            with prediction of Eq.(\ref{eq:eq7}) for parameters $\Phi_{N_P=A} = 0.48$ MeV
            and $b = 4.8$ MeV.}
   \label{Fig:4}
  \end{center}
  \vspace{-0.5cm}
\end{figure}

Finally, we note that the transverse-momentum fluctuations
measured in nuclear collisions at $158$A GeV \cite{NA49pT} and
quantified by the measure $\Phi\left(p_T\right)$ \cite{FOOT10}
show a similar centrality dependence, namely, as seen in Fig.
\ref{Fig:4},
\begin{equation}
\Phi \left( p_T\right) = \Phi_{N_P = A} + b \left( 1 -
\frac{N_P}{A}\right). \label{eq:eq7}
\end{equation}
This behavior of the transverse-momentum fluctuations as a
function of collision centrality was related in superposition
model to the centrality dependence of the multiplicity
fluctuations \cite{MRW} using the correlation of average
transverse momentum and multiplicity observed in collisions
\cite{NA49pT}. To derive Eq. (\ref{eq:eq7}) notice that in general
$\Phi$ is governed by the multiplicity fluctuations in the
following way \cite{RWW}:
\begin{widetext}
\begin{equation}
\Phi\left( p_T\right) = \sqrt{ Var\left( p_T\right) + 2\langle
p_T\rangle^2\frac{Var(N)}{\langle N\rangle}\left[ 1 - \rho
\sqrt{\frac{Var\left( p_T\right)}{\langle
p_T\rangle^2}\frac{\langle N\rangle}{Var(N)} + 1}\right] } -
\sqrt{ Var\left( p_T\right) }, \label{eq:eq5}
\end{equation}
\end{widetext}
where $\rho$ is the correlation coefficient between $N$ and
$\sum^N_i p_{iT}$. As in \cite{MRW} we can see that multiplicity
fluctuations determine the behavior of $\Phi\left( p_T\right)$. In
the first order approximation and taking into account multiplicity
fluctuations given by Eq.(\ref{eq:eq4}) we can write the
transverse momentum fluctuations measure as
\begin{eqnarray}
\Phi\left( p_T\right) &\simeq& \sqrt{ Var\left( p_T\right) }\left[
\omega_0 ( 1 - \rho) \frac{ \langle p_T\rangle^2}{ Var\left(
p_T\right)} - \frac{1}{2}\rho \right] +\nonumber\\
+&& \!\!\! \!\!\!\!\!\!( 1 - \rho)\sqrt{ Var\left( p_T\right)}
\frac{\langle p_T\rangle^2}{ Var\left(
p_T\right)}\frac{1}{a'}\left( 1 - \frac{N_P}{A}\right),
\label{eq:eq6}
\end{eqnarray}
which has the form of Eq. (\ref{eq:eq7}) in what concerns
dependence on the variable $N_P/A$. The slope parameter $b$
corresponds to the correlation coefficient $\rho \sim 0.99$ (for
the previously estimated value of $a'$ and for the transverse
momentum fluctuations $Var(p_T)/\langle p_T\rangle^2 \simeq 0.43$,
as observed experimentally \cite{CERES}). Once again we observe
scaling behavior in the variable $f = N_P/A$.

\section{\label{sec:IV}Summary}

We have investigated some features of the multiplicity
fluctuations observed in recent high-energy nuclear collisions by
attributing them to the intrinsic, nonstatistical, fluctuations of
the temperature of the hadronizing system produced in such
collisions. To this end we used the Tsallis statistics approach in
which such fluctuations are accounted for by the nonextensivity
parameter $q$ (more precisely by $q-1$). When fluctuations are
vanishing $q \rightarrow 1$, Tsallis statistics becomes the usual
Boltzmann-Gibbs (BG) one and the power-like Tsallis $q$-exponents
become the usual exponential distributions. We were considering a
generalized $q$-exponential ensemble describing system of $N_P$
participating nucleons (assumed to be proportional to the size of
hadronizing system). It was found that in this case one can
associate the nonextensivity parameter $q$ with the number of
participants $N_P$ (cf. Eq.(\ref{eq:q-1})). It was found that data
require a generalization of the usual notion of fluctuating
temperature by adding effects of the energy transfer between the
hadronizing source and the surroundings composed with nucleons not
participating directly in the reaction. This resulted in the
introduction of a $q$-dependent effective temperature $T_{eff}$.
We have also allowed for fluctuations of the full accessible
energy $U$ (cf., Section \ref{sec:IIIA}), without these
fluctuations $Var(N)/\langle N\rangle$ would be simply constant,
independent of $N_P/A$.

The simplest possible explanation of the observed effect was the
$q-1$ dependence on the number of participants $N_P$ provided by
the Eq.(\ref{eq:q-1}). We assumed that parameter connected with
the size of the collision region, $\xi$, is approximately given by
$\xi = N_P/A$. The modifications of this assumption are possible
and lead to better agreement with experimental data. Among others,
the presence of nonlinear terms in the $\xi$ variable in
Eq.(\ref{eq:relation})  would approve consistency with data (or,
for example, assuming that $\xi = \zeta (N_P/A)$ with $\zeta
(N_P/A)$ more complicated as used here where $\zeta (N_P/A) =
N_P/A$). Nevertheless, we restricted our discussion to the
simplest possible approximation to demonstrate the connection
between temperature fluctuations and observables from relativistic
ion collisions in a more transparent way.

Let us close with some remarks concerning limitations of our
analysis. Firstly, let us first remind that multiplicity
fluctuations considered in this study were observed for a fixed
number of the projectile participants. Although, in the collisions
of identical nuclei, the average numbers of the participants from
the projectile equal approximately that from the target, the
actual number of target participants fluctuate whereas the number
of the projectile participants is kept fixed. However, one should
notice that the multiplicity fluctuations were measured in the
forward hemisphere ($1.1 < y_{c.m.} < 2.6$) in which the influence
from the target participants is marginal. The models which
produce, independently of centrality, approximately Poissonian
multiplicity distributions highly underestimate the observed
multiplicity fluctuations in noncentral collisions. They are
unable to reproduce, even qualitatively, the centrality dependence
of the scaled variance (cf., for example, \cite{NA49} and
references therein). The highest scaled variance is obtained from
the hadron-string dynamics \cite{KLGB} and ultrarelativistic
molecular dynamics \cite{UrQMD}. However, both approaches (in
their standard versions) show a flat scaled variance, $\omega
\simeq 1.2$, and exhibit almost no dependence on $N_p$
\cite{MIG4}. Finally, note that in comparison with experimental
data (which have finite acceptance, equal to $p=0.16$ in the case
of NA49 data \cite{NA49}) we have used multiplicity $N$ given by
this acceptance. Also all parameters used for description of these
data (like $n_0$, $a'$, $\omega_0$, $\Phi_{N_p=A}$ and $b$) are
acceptance dependent, However, as shown in \cite{FOOT-I}, the
nonextensivity parameter $q$ and its dependence on $N_p$ obtained
from comparison with experimental data does nor depend on
experimental acceptance.

\begin{acknowledgments}
Partial support (GW) of the Ministry of Science and Higher
Education under contracts 1P03B02230 is acknowledged.
\end{acknowledgments}

\appendix

\section{\label{sec:Teff_derivation}Temperature fluctuations}

We shall collect here the main points concerning the idea of
temperature fluctuations \cite{WW,WW_epja}.

Suppose that we have a nonhomogeneous thermodynamic system in
different regions of which there is different temperature $T$,
which fluctuates around some mean temperature $T_0$. As result of
these fluctuations, the actual temperature $T$ equals
\begin{equation}
T = T_0 - \xi(t) T , \label{eq:Tfluct1}
\end{equation}
where $\xi (t)$ describes the actual (not specified) stochastic
process causing these fluctuations.

There must take place an exchange of energy (heat) between the
regions mentioned above, in particular between any selected region
and the rest of the system. This exchange eventually leads to
equilibration of the temperature of the whole system. The
corresponding process of heat conductance is described by the
following equation \cite{LL}:
\begin{equation}
c_p\rho \frac{\partial T}{\partial t} - \gamma (T' - T) = \phi ,
\label{eq:Tfluct2}
\end{equation}
where $c_p$, $\rho$ and $\gamma$ are, respectively, the specific
heat under constant pressure, the density and the coefficient of
external conductance. The heat source term $\phi$ determines the
amount of  energy transfer  per unit time and unit volume. Using
Eqs. (\ref{eq:Tfluct1}) and (\ref{eq:Tfluct2}) one gets the
following Langevine equation for the temperature $T$:
\begin{equation}
\frac{\partial T}{\partial t} + \left[ \frac{1}{\tau} +
\xi(t)\right]T = \frac{1}{\tau} \left[ T_0 +
\frac{\tau}{c_p\rho}\phi \right] \label{eq:Tfluct3}
\end{equation}
with coefficient $\tau = c_p\rho/\gamma$. As stated before, $\xi
(t)$ describes stochastic changes of temperature in time. Let us
assume that these changes are such that their mean value is zero,
\begin{equation}
\langle \xi(t)\rangle = 0, \label{eq:zeroaverage}
\end{equation}
whereas, for sufficiently fast changes, its correlator is equal to
\begin{equation}
\langle \xi(y) \xi(t + \Delta t)\rangle = 2 D \delta(\Delta t).
\label{eq:correlator}
\end{equation}
The constants $\tau$ and $D$ define, respectively, the
characteristic mean time for the temperature changes and their
variance:
\begin{eqnarray}
\langle T(t)\rangle &=& T_0 + T(t=0)\exp\left( - \frac{t}{\tau}
\right), \label{eq:mean}\\
\langle T^2(t = \infty)\rangle &=& \frac{1}{2} D \tau.
\label{eq:variance}
\end{eqnarray}
Thermodynamic equilibrium in such a situation means that for $t
>> \tau$ the influence of the initial condition $T(t=0)$ vanishes
and the mean squared $T$ has a value corresponding to the state of
equilibrium.

Eq. (\ref{eq:Tfluct3}) leads to the corresponding Fokker-Planck
equation,
\begin{equation}
\frac{d f(T)}{dt} = - \frac{\partial}{\partial T} K_1 f(T) +
\frac{1}{2} \frac{\partial ^2}{\partial T^2} K_2 f(T),
\label{eq:FP}
\end{equation}
where the intensity coefficients are
\begin{eqnarray}
K_1(T) &=& \frac{1}{\tau}\left( T_0 + \frac{\tau}{c_p\rho}\phi
\right)  + \left(D- 2\frac{1}{\tau}\right) T,\nonumber\\
K_2(T) &=& 2 DT^2. \label{eq:K1K2}
\end{eqnarray}
Its solution is gamma distribution in $1/T = \beta$:
\begin{equation}
f(T) = \frac{\mu}{\Gamma
(\alpha)}\left(\frac{\mu}{T}\right)^{\alpha - 1} \exp\left( -
\frac{\mu}{T}\right), \label{eq:Gamma}
\end{equation}
where
\begin{equation}
\alpha = \frac{1}{\tau D},\quad \mu = \frac{1}{\tau D}\left(T_0 +
\frac{\tau}{c_p\rho}\phi\right). \label{eq:alphamu}
\end{equation}
The mean value and the relative variance of $\beta$ are,
respectively \cite{FOOT11},
\begin{eqnarray}
&& \langle \beta\rangle = \left( T_0 - \frac{\tau}{c_p\rho}\phi
\right)^{-1},\label{eq:mean1}\\
&&\frac{\langle \beta^2\rangle - \langle \beta\rangle^2}{\langle
\beta\rangle^2} = \frac{1}{\alpha} = \tau D. \label{eq:var}
\end{eqnarray}
Temperature fluctuations in the form presented here when applied
to the exponential Boltzmann-Gibbs formula, Eq. (\ref{eq:exp}),
lead to $h_q(E) = \int_0^{\infty} \exp(-E/T)f(T)d(1/T) $, i.e., to
a Tsallis distribution, Eq. (\ref{eq:q_exp}),  with the
nonextensivity parameter $q$ equal to \cite{WW}
\begin{equation}
q = 1 + \frac{1}{\alpha} = 1 + \tau D \label{eq:defq1}
\end{equation}
and with the effective temperature
\begin{equation}
T_{eff} = T_0 + \frac{\tau}{c_p\rho}\phi = T_0 + (q -
1)\frac{1}{Dc_p\rho}\phi . \label{eq:defTeff}
\end{equation}
In the case of no energy transfer, i.e., when $\phi =0$, one is
left only with fluctuations and then $T_{eff} = T_0 $, as in
\cite{WW}. Recently temperature fluctuations due to the volume
fluctuations in statistical models were discussed as well and
introduced in modelling of relativistic particle collision
processes \cite{BGG}.

\section{\label{sec:P(N)_derivation}Multiplicity distributions in
Boltzmann and Tsallis ensembles}

We shall now recapitulate some basic ideas concerning multiplicity
distributions which result from, respectively, Boltzmann and
Tsallis ensembles \cite{WWfluct}.

\subsection{\label{sec:Poisson}Poisson multiplicity distribution}

This distribution arises in situations where, in some process, one
has $N$ independently produced secondaries with energies
$\{E_{1,\dots,N}\}$ distributed according  to a Boltzmann
distribution:
\begin{equation}
g(E_i) = \frac{1}{\lambda}\exp \left( -
\frac{E_i}{\lambda}\right);\quad {\rm where}\quad \lambda =
\langle E\rangle. \label{eq:B1}
\end{equation}
The corresponding joint probability distribution is given by
\begin{equation}
g\left( \{ E_{1,\dots,N} \} \right) = \frac{1}{\lambda ^N} \exp
\left( - \frac{1}{\lambda} \sum^N_{i=1} E_i \right). \label{eq:B2}
\end{equation}
For independent energies $\{E_{1,\dots,N}\}$ the sum $E =
\sum^N_{i = 1} E_i$  is distributed according to a gamma
distribution,
\begin{equation}
g_N(E) = \frac{1}{\lambda(N - 1)!}\left(
\frac{E}{\lambda}\right)^{N-1} \exp \left( - \frac{E}{\lambda}
\right), \label{eq:B3}
\end{equation}
the distribuant of which is
\begin{equation}
G_N(E) = 1 - \sum^{N-1}_{i=1}\frac{1}{(i - 1)!}\left(
\frac{E}{\lambda}\right)^{i-1} \exp \left( - \frac{E}{\lambda}
\right). \label{eq:B4}
\end{equation}
Eq. (\ref{eq:B3}) follows immediately either by using
characteristic functions or by sequentially performing integration
of the joint distribution (\ref{eq:B2}) and noticing that
\begin{equation}
g_N(E) = g_{N-1}(E)\frac{E}{N-1}. \label{eq:B5}
\end{equation}
For energies such that
\begin{equation}
\sum^N_{i=0}E_i \leq E \leq \sum^{N+1}_{i=0} E_i, \label{eq:B6}
\end{equation}
the corresponding multiplicity distribution has a Poissonian form
(with $\langle N\rangle = \frac{E}{\lambda}$)
\begin{equation}
P(N) = G_{N+1}(E) - G_N(E) = \frac{\langle N\rangle^N}{N!} \exp( -
\langle N\rangle). \label{eq:B7}
\end{equation}
Therefore, whenever $N$ variables $\{ E_{1,\dots,N,N+1}\}$ follow
an exponential distribution (\ref{eq:B1}) and satisfy condition
(\ref{eq:B6}), then the corresponding multiplicity $N$ has a
Poissonian distribution (\ref{eq:B7}).

\subsection{\label{sec:NBD}Negative Binomial multiplicity
distribution}

This distribution arises when in some process $N$ independent
particles with energies $\{E_{1,\dots,N}\}$  are distributed
according to a Tsallis distribution,
\begin{equation}
h\left( \{E_{1,\dots,N}\}\right) = C_N \left[ 1 - (1 - q)
\frac{\sum^N_{i=1}E_i}{\lambda} \right]^{\frac{1}{1-q} + 1 - N}.
\label{eq:B8}
\end{equation}
It means that there are some intrinsic (so far unspecified but
summarily characterized by the parameter $q$) fluctuations present
in the system under consideration. In this case we do not know the
characteristic function for the Tsallis distribution. However,
because we are dealing here only with variables
$\{E_{1,\dots,N}\}$ occurring in the form of the sum, $E =
\sum^N_{i=1} E_i$ , one can still sequentially perform
integrations of the joint probability distribution (\ref{eq:B8})
arriving at the formula corresponding to Eq. (\ref{eq:B3}):
\begin{eqnarray}
h_N(E) &=& h_{N-1}(E)\frac{E}{N-1} =
\frac{E^{N-1}}{(N-1)!\lambda^N}\cdot
\label{eq:B10}\\
\cdot &&\!\!\!\!\!\!\!\!\!\! \prod^N_{i=1}[(i - 2)q - (i-3)]
\left[ 1 - (1 - q)\frac{E}{\lambda}\right]^{\frac{1}{1-q} + 1 -
N}\nonumber
\end{eqnarray}
with distribuant given by
\begin{equation}
H_N(E) = 1 - \sum^{N-1}_{j=1} \tilde{H}_j(E)
\label{eq:HE}
\end{equation}
where
\begin{eqnarray}
\tilde{H}_j(E) &=&
\frac{E^{j-1}}{(j-1)!\lambda^j}\cdot \label{eq:B12}\\
\cdot &&\!\!\!\!\!\!\!\!\!\! \prod^j_{i=1}[(i - 2)q - (i-3)]\left[
1 - (1 - q)\frac{E}{\lambda}\right]^{\frac{1}{1-q} + 1 -
j}\nonumber.
\end{eqnarray}
As before, for energies $E$ satisfying the condition given by
Eq.(\ref{eq:B6}), the corresponding multiplicity distribution is
equal to the well known Negative Binomial distribution (NBD):
\begin{eqnarray}
P(N) &=& H_{N+1}(E) - H_N(E) = \label{eq:B13}\\
&=& \frac{(q-1)^N}{N!}\frac{q-1}{2-q}\frac{\Gamma\left( N + 1 +
\frac{2-q}{q-1}\right)}{\Gamma \left(
\frac{2-q}{q-1}\right)}\cdot\nonumber\\
&&\cdot  \left( \frac{E}{\lambda}\right)^N\left[ 1 - (1
-q)\frac{E}{\lambda}\right]^{-N + \frac{1}{1-q}} =
\nonumber\\
&=&\frac{\Gamma(N + k)}{\Gamma(N + 1)\Gamma(k)}\frac{\left(
\frac{\langle N\rangle}{k}\right)^N}{\left( 1 + \frac{\langle
N\rangle}{k}\right)^{N + k}} \label{eq:B14}
\end{eqnarray}
where
\begin{eqnarray}
k &=& \frac{1}{q-1},\qquad \quad \langle N\rangle =
\frac{E}{\lambda},\nonumber\\
Var(N) &=& \frac{E}{\lambda}\left[ 1 - (1 -
q)\frac{E}{\lambda}\right] \nonumber\\
&=& \langle N\rangle + (q - 1) \langle N\rangle^2. \label{eq:B15}
\end{eqnarray}
In cases considered here $q \in (1,2)$. In the limiting cases of
$q \rightarrow 1 $ one has $k \rightarrow \infty$ and $P(N)$
becomes a Poisson distribution whereas for $q \rightarrow 2$ one
has $ k \rightarrow 1$ and $P(N)$ becomes the so called
geometrical distribution.



\begin{thebibliography}{99}
\bibitem{FOOT3} In a sense, our work is a continuation of many previous works
                \cite{qWW,qNUWW,NUWW,C_V_Q} showing that in order to adequately
                describe experimental data on energy and transverse momenta
                distributions the usual statistical approach based on Boltzmann
                distributions should be modified and replaced by a generalized
                nonextensive formalism based on Tsallis statistics \cite{Tsallis}.
\bibitem{FOOT4} Actually, in the work \cite{WW_epja} accounting for the effect
                of viscosity, the source term in (\ref{eq:Teff}) takes form
                $\phi = \eta f(u)$, where $\eta$ denotes coefficient of viscosity
                and function $f(u)$ contains terms dependent on the velocity $u$
                in the form of $\partial u_i/\partial x_k + \partial u_k/\partial x_i$
                (as in \cite{LL}).Therefore the effective temperature there was given by
                $T_{eff} = T_0 + (q-1)\eta f(u)/(Dc_p\rho)$.
\bibitem{FOOT5} It should be noted that in \cite{DataPi} the $p_T$ spectra were
                analyzed in terms of the so called escort probability distributions
                \cite{escort}, i.e.,
                $h_Q(E) = C_Q\left[ 1 - (1 - Q)E/\Lambda\right]^{Q/(1-Q)}$.
                However, this distribution function is, in fact, formally
                identical with
                $h_q(E) = C_q\left[ 1 - (1 - q)E/\lambda\right]^{1/(1-q)}$,
                which we are using in this work, provided that we identify:
                $q = 1 (Q-1)/Q$, $\lambda = \Lambda/Q$ and
                $C_q = C_Q = (2-q)/\lambda = 1/\Lambda$.
                The mean value is now
                $\langle E\rangle = \lambda/(3-2q) = \Lambda/(2-Q)$
                and is defined for $q \in (0, 1.5)$ (to be compared with
                previous $Q \in (1/2,2)$).
\bibitem{FOOT6} Recently Tsallis statistics was implemented in the so called
                Blast-Wave model and applied to the transverse momentum spectra
                measured at RHIC \cite{BWM}. As a result the dependence of
                temperature and collective flow velocity on the $(q-1)$ parameter
                has been advocated. Similar dependence was also found in a recent
                analysis of the non-thermal equilibrium in heavy-ion collisions
                performed using Tsallis distributions \cite{CHLW}.
\bibitem{FOOT-I} This is because the mean accepted multiplicity for
                 acceptance $p$ is $\langle N\rangle = p\langle N_{p=1}\rangle$
                 and the scaled variance for the accepted particles is
                 $Var(N)/\langle N\rangle = (1 - p) +
                 pVar\left( N_{p=1} \right)/\langle N_{p=1}\rangle$,
                 therefore one gets from the Eq. (\ref{eq:der1}) that
                 nonextensivity parameter obtained this way does not
                 depend on acceptance, i.e., $q = q_{p=1}$.
\bibitem{FOOTA} The fitted value of $n_0 = 0.642$ is only a little greater than the
                multiplicity observed in $p+p$ collisions when calculated using the
                acceptance of the NA49 experiment. Notice also that value of
                $c = 4.1$ obtained here for $Pb+Pb$ collisions is not far from the
                value $1.25/0.22 = 5.7$ obtained for data from RHIC (i.e., for
                Au+Au collisions but at much higher energy) which we have obtained in
                Fig. \ref{Fig:1}.
\bibitem{FOOT7} One should keep in mind that both the $q$ and $T_{eff}$ evaluated
                from transverse characteristics of the reaction can differ from those
                obtained from its longitudinal characteristics. It means that
                these parameters in multiplicity distributions, $P(N)$ (which are
                sensitive to $p=\sqrt{p_L^2 + p_T^2}$) can differ substantially
                from those obtained by analyzing transverse momentum distribution.
                Actually, in \cite{NUWW} we advocate that $|1 - q_T| << |1 -
                q_L|$. Assuming now that $Var(T) = Var(T_T) + Var(T_L)$ we obtain
                that $q \langle T\rangle ^2 = q_L \langle T_L\rangle^2 + q_T
                \langle T_T\rangle ^2 + \langle T\rangle ^2 - \langle T_L\rangle
                ^2 - \langle T_T\rangle ^2$, which for $\langle T_L\rangle >>
                \langle T_T\rangle$, as observed in \cite{qNUWW}, leads to $q
                \simeq q_L$.
\bibitem{FOOT8} In the same way as the (improper)eigenstates of position and momentum
                appear as extreme cases in the quantum mechanical uncertainty relations.
                It is worth knowing that in \cite{limq} the limiting cases of Tsallis
                statistics was investigated in which $q$ was interpreted as a measure
                thermal bath heat capacity: $q = 1$, i.e. canonical, case would
                correspond to an infinite bath (thermalized and with fixed
                temperature), whereas $ q = - \infty $, i.e. microcanonical, case
                would correspond to a bath with null heat capacity (isolated and
                with fixed energy). All intermediate cases would then correspond
                to the finite heat capacity (both temperature and energy fluctuate).
\bibitem{FOOT9} In our formalism presented here we have (cf. Eq. (\ref{eq:B3})) that
                $g_{T,N}(U) = [\beta/\Gamma(N)](\beta U)^{N-1}\exp( - \beta U)$
                with fluctuations given by $Var(U)/\langle U\rangle^2 = 1/N$. One
                can now invert distribution $g_{T,N}(U)$ proceeding in analogous
                way as used in Appendix \ref{sec:P(N)_derivation} to obtain
                multiplicity distributions $g_{T,U}(N)$ and obtain distribution of
                temperature, $g_{U,N}(T) = \frac{\partial}{\partial \beta}\int_0^U
                g_{T,N}(U') dU' = [U/\Gamma(N)](\beta U)^{N-1} \exp(-\beta U)$
                with fluctuations $Var(T)/\langle T\rangle ^2 \simeq
                Var(\beta)/\langle \beta \rangle ^2 = 1/N  $. The fluctuations of
                temperature obtained this way have gamma distribution analogous to
                distribution (\ref{eq:Gamma}). Because $C_V \propto N$ one obtains
                Eqs. (\ref{eq:eq1}) and (\ref{eq:microcanonical}). This
                illustrates that we can deduce fluctuations (i.e., the
                corresponding probability distributions) of any quantity out of
                ($T, U, N$) provided only that the other two are constant. The
                implication of this fact will be discussed elsewhere.
\bibitem{FOOT10} Measure $\Phi\left( p_T\right)$ has been
introduced in \cite{PHI} and is defined as $\Phi = \sqrt{\langle
Z^2\rangle/\langle N\rangle} - \sqrt{\langle z^2\rangle}$ where $z
= p_T - \langle p_T\rangle$ and $Z = \sum^N_i z_i$.
\bibitem{FOOT11} Notice that the relative variances of temperature $T$ and its
inverse $\beta$ are roughly the same, $\frac{\langle
\beta^2\rangle - \langle \beta\rangle^2}{\langle \beta\rangle^2}
\simeq \frac{\langle T^2\rangle - \langle T\rangle^2}{\langle
T\rangle^2}$.
\end{thebibliography}
\end{document}